\documentclass[12pt]{article}

\usepackage{amsfonts,latexsym,fullpage,amsmath,latexsym,amssymb}

\newtheorem{Definition}{Definition}
\newtheorem{Theorem}{Theorem}
\newtheorem{Lemma}{Lemma}

\newtheorem{Observation}{Observation}
\newtheorem{example}{Example}

\newcommand{\onemat}[0]{{\mathbf 1}}
\newcommand{\cM}[0]{{\mathcal M}}
\newcommand{\cT}[0]{{\mathcal T}}
\newcommand{\C}[0]{{\mathbb{C}}}
\newcommand{\F}[0]{{\mathbb{F}}}
\newcommand{\R}[0]{{\mathbb{R}}}
\newcommand{\N}[0]{{\mathbb{N}}}

\newcommand{\SL}[0]{{\rm SL}}
\newcommand{\GL}[0]{{\rm GL}}
\newcommand{\DFT}[0]{{\rm DFT}}
\newcommand{\qed}[0]{\hfill $\Box$}

\begin{document}

\title{\Large \textbf{Universal Simulation of Hamiltonians\\ 
 Using a Finite Set of Control Operations}}
\author{
Pawe{\l} Wocjan\thanks{e-mail: {\protect\tt $\{$wocjan,roettele,janzing,eiss\_office$\}$@ira.uka.de}}, 
Martin R{\"o}tteler, Dominik Janzing, Thomas Beth \\
\small Institut f{\"u}r Algorithmen und Kognitive Systeme,
Universit{\"a}t Karlsruhe,\\[-.5ex]
\small  Am Fasanengarten 5, D-76\,128 Karlsruhe, Germany\\[-.5ex]
\small Forschungsgruppe Quantum Computing}
\date{September 13, 2001}
\maketitle

\abstract{Any quantum system with a non-trivial Hamiltonian is able to
simulate any other Hamiltonian evolution provided that a sufficiently
large group of unitary control operations is available. We show that
there exist finite groups with this property and present a sufficient
condition in terms of group characters. We give examples of such
groups in dimension $2$ and $3$. Furthermore, we show that it is
possible to simulate an arbitrary bipartite interaction by a given one
using such groups acting locally on the subsystems.}

%
%

\section{Introduction}
Simulating Hamiltonian evolutions of arbitrary quantum systems
is an interesting application for future quantum computers.
Historically, the idea of simulating Hamiltonian time evolutions was
the first motivation for quantum computation \cite{feynmann}. Whereas
in early works on this problem the desired Hamiltonian was proposed to
be simulated by a discrete sequence of gate operations (see
e.\,g.~\cite{Lloyd96}), a more control theoretic formulation of the
problem has become popular recently
\cite{Khaneja01,dodd,graph,two,stoll,vidal}. In this formulation one
assumes that the dynamics of the quantum computer is determined by its
Hamiltonian together with external control possibilities. Here the
task is to simulate an evolution that would have occurred under some
other Hamiltonian by interspersing the natural time evolution with
control operations. More explicitly, one assumes that the natural
evolution $\exp(-iHt)$ alternates with fast implementations of unitary
operations $V_j$, i.\,e., the resulting evolution is given by
\begin{equation}\label{resultingEvolution}
\exp(-i H \tau_N t) V_N \cdots \exp(-i H \tau_2t) V_2 \exp(-i H \tau_1 t) V_1\,,
\end{equation}
where the relative times between the control operations are given by
$\tau_i$ for $i=1,\ldots,N$. If the $t$ is small compared to the
time scale of the evolution $\exp(-i H t)$, the resulting evolution is
approximatively given by
\[
\exp(-i \bar{H} t)\,.
\]
Here $\bar{H}$ is the {\it average Hamiltonian}
\[
\bar{H}:=\sum_{i=1}^N \tau_i U_i^\dagger H U_i\,,
\]
where we set $U_i:=\prod_{j=1}^i V_j$. Important characteristics of
simulation schemes like (\ref{resultingEvolution}) are the amount of
operations performed (complexity) and the total operation time
(overhead). The average Hamiltonian method is the basis for simulating
Hamiltonians by a given one and has applications which go beyond the
usual aims of quantum computation. As examples we mention the
following applications:

\begin{enumerate}
\item
{\bf Decoupling / Suppression of Decoherence}

The time evolution on the joint Hilbert space of a system and its
environment can be interspersed with transformations on the system's
space alone in such a way that the net effect is a separate
(``decoupled'') time evolution of the system and the bath
\cite{Zanardi00}. This is a generalization of decoupling techniques in
Nuclear Magnetic Resonance \cite{ernst}.

\item
{\bf Time inversion}

The natural time evolution $t\mapsto \exp(-iHt)$ with $t\geq 0$ can be
conjugated by unitary transformations in such a way that the total
effect is a transformation of the form $\exp(-iHs)$ with negative
$s$. Remarkably, there are schemes for inverting {\it unknown}
Hamiltonians. This fact is closely related to the existence of
decoupling schemes \cite{Zanardi00}. Time inversion for unknown
Hamiltonians is a useful primitive in {\it quantum process
tomography}, i.\,e., procedures that distinguish between unknown time
evolutions \cite{JBdistinguish}. The essential idea is that time inversion
enables to implement transformations of the form $\exp([H,A])$ for
arbitrary self-adjoint $A$ even if $H$ is unknown.

\item
{\bf Generating arbitrary time evolutions with a finite control group}

Assume that the only external control operations of a quantum system
are implementations of unitary operations taken from a finite group
$S$. If the natural time evolution is non-trivial and $S$ in a sense
is large enough, than every unitary transformation can be generated by
concatenations of the natural evolution and elements of $S$. In
particular, this is possible if the natural Hamiltonian $H$ can be
transformed into any other in the sense of the average Hamiltonian
method.

This is a special instance of the following problem: Given an
arbitrary $\R$-linear map $L$ on the set of self-adjoint traceless
operators, the task is to find a sequence of unitary operations such
that the system evolves according to the Hamiltonian $L(H)$ if its
natural (unknown) Hamiltonian $H$ is present. The problems of
inverting or switching off unknown Hamiltonians are special cases of
this problem with $L=0$ or $L(H)=-H$. In its full generality,
procedures for simulating $L(H)$ if an unknown $H$ is present can be
used as primitives in quantum process tomography \cite{JBdistinguish}
and quantum control procedures \cite{JZAB01}.
\item
{\bf Simulating interactions by other ones}

Consider a bipartite system with Hilbert space
$\C^d\otimes\C^d$. Assume that the interaction between both parts is a
fixed Hamiltonian on the joint Hilbert space which cannot be
controlled at all. The only possibilities to control the system are
implementations of local transformations on each of the
subsystems. These transformations can be used to imitate other
interactions \cite{two,vidal}.

In $n$-partite systems $(\C^d)^{\otimes n}$, where the interaction
between each pair of subsystems is assumed to be fixed, one can
simulate other pair-interactions in the sense of the average
Hamiltonian method. Time optimization of these simulations leads to
interesting problems of parallel execution. For $n$-qubit networks,
this problem has been addressed in \cite{graph,arrow,stoll,leung}.
\end{enumerate}
The paper is organized as follows. In Section~\ref{dynamical} we
define what it means to simulate a Hamiltonian and explain the
physical meaning in the context of average Hamiltonian theory
\cite{ernst,haeberlen,slichter}. 

In Section~\ref{annihilator} we introduce the concept of an
\emph{annihilator} for $d$-dimensional quantum systems characterizing
control procedures that switch off the possibly unknown dynamics of
the system. These procedures directly give decoupling and inversion
schemes. Explicitly, they can be constructed using nice error bases
\cite{knill96,KR:2000a} which yield minimal annihilator procedures
attaining the lower bound $d^2$ on the complexity. We show a lower
bound on the \emph{overhead} of inversion procedures of a general,
possibly unknown Hamiltonian to be $d-1$ and an upper bound of
$d^2-1$. Furthermore, a lower bound on the complexity is shown to be
$d^2-1$.

In Section~\ref{transformer} we address the question how to simulate an
arbitrary Hamiltonian by any other if only a restricted set of control
operations is available. The condition on the set of available control
operations for making universal simulation possible is stronger than
the requirement to make annihilation possible. A sufficient condition
for a group of available control operations allowing universal
simulation can be formulated in terms of group characters. Groups
satisfying this criterion will be called {\it transformer groups}.  We
present transformer groups for dimensions 2 and 3. Furthermore, we
show in Section~\ref{bipartite} that the transformer groups allow to
transform any interaction in a bipartite system into any other by
operating on the subsystems only.

%
%

\section{Dynamical control}\label{dynamical}
For the implementation of a quantum computer it is necessary to
control the time evolution of the used physical system in a universal
way.  In many physical systems the only directly accessible control
possibilities are given by a set of control unitaries and the system
Hamiltonian that cannot be switched off. By applying the control
operations we effectively change the Hamiltonian into a piecewise
constant time-dependent Hamiltonian. The formalism of average
Hamiltonian theory \cite{ernst,haeberlen,slichter} allows to solve for
the resulting time evolution at a time $t$ by writing the evolution of
a time independent average Hamiltonian $\bar{H}$.  Following
\cite{ernst,haeberlen,slichter} we briefly sketch average Hamiltonian
theory: the overall dynamic after a period of evolution is
given by
$$ U(t)=\mathcal{T} \exp(-i\int_0^t\mathrm{d}\tau H(\tau)) =
\exp(-i\bar{H} t)\,, 
$$ 
where $\mathcal{T}$ denotes the Dyson time ordering operator. A
solution of this equation is a time independent Hamiltonian that would
result in the same unitary if it were applied over the same period. If
the Hamiltonian $H(\tau)$ commutes with itself at all times we have
$\bar{H}=\int_0^t \mathrm{d}\tau H(\tau)$. However, this is rarely the
case. For sufficiently small $t$, the Magnus expansion provides a
formal means of calculating the average Hamiltonian:
\begin{equation}
  \bar{H}=\bar{H}^{(0)}+\bar{H}^{(1)}+\bar{H}^{(2)}+\ldots
\end{equation}
where the operators $\bar{H}^{(0)},\bar{H}^{(1)},\ldots$ are the
average Hamiltonians of increasing order
\begin{eqnarray}
  \bar{H}^{(0)} & = & \frac{1}{t}\int_0^t\mathrm{d}{\tau} H(\tau), \\
  \bar{H}^{(1)} & = & \frac{-i}{2t}\int_0^t\mathrm{d}\tau'
    \int_0^t\mathrm{d}\tau''
    [H(\tau'),H(\tau'')]\,.
\end{eqnarray}
We have $\|H^{(0)}\|\le 1$ and $\|\bar{H}^{(1)}\|\le t/2$ since
$\|[H(\tau'),H(\tau'')]\|\le 2\|H(\tau')\|\|H(\tau'')\|=2$ and we are
integrating over the simplex of area $t^2/2$. The norm of the higher
order terms is bounded by higher orders of $t$. Therefore for
sufficiently small time $t$ the resulting unitary $U(t)$ is
essentially determined by $\bar{H}^{(0)}$.

We assume that the only directly accessible control possibilities are
the unitaries in the \emph{control} set ${\mathcal C}\subseteq SU(d)$
and assume in addition that they can be performed arbitrarily fast
compared to the natural time evolution. This socalled \emph{fast
control limit} is justified e.\,g.\ in NMR because the coupled and
local evolutions act on significantly different time scales. We will
now subject the system to a cyclic pulse train. The pulses are assumed
to be infinitely short. A sequence consisting of $N$ pulses will be
denoted by
\begin{equation}
  P:=(V_1,\tau_1,\ldots,V_N,\tau_N)
\end{equation}
where $V_i\in {\mathcal C}$ and $\tau_i>0$ are relative times, i.\,e.\
$\tau_i>0$ and $\sum_{i=1}^N \tau_i=1$. The pulses are applied from
left to right. If we apply the sequence over the time $t$ the
resulting unitary is given by 
\begin{equation}
U_P(t)=\prod_{i=1}^N \exp(-i H\tau_i t) V_i\,.
\end{equation}
The $\tau_i$ specify the fraction of time between the pulses. For a
cyclic sequence (defined by $\prod_{i=1}^N V_i=\onemat$) we can express the
resulting unitary as
\begin{equation}
U_P(t)=\prod_{i=1}^N U_i^{\dagger} \exp(-i H \tau_i t) U_i\,,
\end{equation} 
where $U_i=\prod_{j=1}^i V_j$. Using the identity
$U^{\dagger}\exp(H)U=\exp(U^{\dagger} H U)$ we get
\begin{equation}
  U_P(t)=\prod_{i=1}^N \exp(-i H_i \tau_i t)\,, 
\end{equation}
where $H_i=U_i^{\dagger} H U_i$. These operators are the Hamiltonians
in the so-called ``toggling frame''. Let $Ad_{\mathcal C}(H)$ denote
set of conjugates of $H$
\begin{equation}
  Ad_{\mathcal C}(H)=\{Ad_U(H)=U^{\dagger} H U\mid U\in {\mathcal
C}\}\,.
\end{equation}
Then the unitary $U_P(t)$ is the solution of a time-dependent
Schr\"odinger equation with piecewise constant Hamiltonians in
$Ad_{\mathcal C}(H)$.

The previous discussion motivates the following definition of the
notion of simulating a Hamiltonian by another Hamiltonian.

\begin{Definition}[First order simulation]\label{simulating}
Let $\tilde{H}$ be any Hamiltonian. We say $\tilde{H}$ can be
simulated by $H$ with overhead $1$, written $\tilde{H}\prec H$, if and
only if there are $\tau_i>0$ summing up to $1$ and $H_i\in
Ad_{\mathcal C}(H)$ such that $\tilde{H}=\sum_{j} \tau_i H_i$, i.\,e.\
$\tilde{H}$ can be written as a convex combination of conjugates of
$H$ by elements of ${\mathcal C}$.  $\tilde{H}$ can be simulated by
$H$ with overhead $\tau$ iff $\tilde{H}\prec \tau H$.
\end{Definition}

Using this definition of simulation the problem of time-optimal
simulation of a Hamiltonian is reduced to a convex optimization
problem (this has been noted in \cite{two,graph}).

%
%

\section{Annihilators}\label{annihilator}
In this section we introduce the concept of an annihilator for a
$d$-dimensional quantum system characterizing control procedures for
switching off the possibly unknown dynamics of the system. These
procedures directly give decoupling and inversion schemes. We prove
some optimality properties of annihilators and show how a minimal
annihilator can be explicitly constructed using nice error basis.
\begin{Definition}[Annihilator]
Let $P:=(V_1,\tau_1,V_2,\tau_2,\ldots,V_N,\tau_N)$ be a cyclic pulse
sequence of length $N$, i.\,e.\ $\tau_i>0$, $\sum_{i=1}^N \tau_i=1$, and
$\prod_{i=1}^N V_i=\onemat$. Set $U_i=\prod_{j=1}^i V_j$. The sequence
$P$ is called an annihilator of dimension $d$ and length $N$ iff
\begin{equation}
\sum_{i=1}^N \tau_i U_i^\dagger a U_i=0
\end{equation}
for all $a\in su(d)$. An annihilator is called minimal if there is no
shorter annihilator.
\end{Definition}

\begin{Theorem}[Minimal annihilator]\label{minimal}
A minimal annihilator has length $d^2$. Furthermore all relative times
$\tau_i$ of a minimal annihilator are equal.
\end{Theorem}
{\bf Proof:} Let $(V_1,\tau_1,V_2,\tau_2,\ldots,V_N,\tau_N)$ be an
arbitrary annihilator.  The corresponding $U_i=\prod_{j=1}^i V_j$ and
$\tau_i$ define a realization of the depolarizing channel $\Lambda$
on ${\mathbb C}^d$ by random external fields \cite{zyczkowski} since
\begin{equation}
\Lambda(\rho)=\sum_{i=1}^N \tau_i U_i^\dagger\rho U_i = \onemat/d \,.
\end{equation}
This permits to show that $N\ge d^2$ as follows: by sending one part
of a maximally entangled state $|\Psi\rangle\langle\Psi |$ in
${\mathbb C}^d\otimes {\mathbb C}^d$ through the depolarizing channel
we end up with the maximally mixed state $\onemat/d^2$. Therefore we
need at least $d^2$ unitaries since the rank of each $(\onemat\otimes
U_i^\dagger)|\Psi\rangle\langle\Psi |(\onemat\otimes U_i)$ is $1$ and
they must sum up to $d^2$ (the rank of the maximally mixed state). For
$N=d^2$ all $\tau_i$ must be equal for entropy reasons (see
\cite{nielsen}, page 518) since for $\tau_i\neq \frac{1}{d^2}$
$$ S\left(\sum_{i=1}^{d^2} \tau_i 
(\onemat\otimes U_i^\dagger)|\Psi\rangle\langle\Psi | 
(\onemat\otimes U_i)\right) 
\le H(\tau_1,\tau_2,\ldots,\tau_{d^2}) < \log_2 d^2 = S(\onemat/d^2)\,, $$
where $S$ denotes here the von-Neumann and $H$ the Shannon entropy. 
One proof of the existence of a minimal annihilator is based on the
concept of nice error basis to be introduced in the next subsection.
\qed

Theorem~\ref{minimal} shows that a minimal annihilator corresponds to
a unitary error depolarizer \cite{werner} and thus to a unitary basis.

\subsection{Nice error bases}
In this section we deal with the problem to construct a minimal
annihilator. For this we construct bases for the vector-space $\C^{d
\times d}$ of $d\times d$-matrices which consist entirely of unitary
matrices and are orthogonal with respect to the trace inner
product. One way of constructing such bases relies on the concept of
\emph{nice error bases}. We refer to \cite{knill96} and
\cite{KR:2000a} for an overview of this method and mention that nice
error bases are used in the construction of quantum error control
codes~\cite{calderbank98,gottesman96,knill96b,steane96}. They are also
of interest in the theory of noiseless
subsystems~\cite{knill00,Zanardi00} and in connection with the
development of quantum authentication codes~\cite{barnum00}.
\begin{Definition}
Let $G$ be a group of order $d^2$ with identity element $e$. A
\emph{nice error basis} on $\mathbb{C}^d$ is a set
$\mathcal{E}=\{U_g\in {\cal U}(d)\mid g\in G\}$ of unitary matrices
such that
\begin{itemize}
\item[(i)]   $U_e$ is the identity matrix,
\item[(ii)]  $\mathrm{tr}\, U_g=d\,\delta_{g,e}$ for all $g\in G$,
\item[(iii)] $U_g U_h=\alpha(g,h)U_{gh}$ for all $g,h\in G$,
\end{itemize}
where the factor system $\alpha(g, h)$ is a function from $G\times G$
to the set $\C^\times := \C \setminus \{0\}$.
\end{Definition}

In \cite{KR:2000a} it was shown that the map $g \mapsto U_g$ defines a
projective representation of $G$; this is a consequence of conditions
(i) and (iii). Condition (ii) shows that the matrices
$U_g$ are pairwise orthogonal with respect to the trace inner product
$\langle A, B \rangle := {\rm tr} (A^\dagger B)/d$. Hence, a nice
error basis is an irreducible projective representation of the finite
group $G$.  The group $G$ itself is also called {\em index group}
since its group elements index the elements of the nice error basis
${\cal E}$.

Note that in general the group generated by the matrices $U_g$ for
$g\in G$ will be larger than $G$, since these matrices are not closed
under multiplication. A well-known theorem from projective
representation theory (cf. \cite[Theorem V.24.6]{HuppertI:83},
\cite[Theorem 11.15]{Isaacs:76}) states that it is always possible to
switch to an equivalent projective representation such that the images
$U_g$ generate a finite group $\hat{G}$ (see also
\cite{KR:2000a}). This group is called the {\em abstract error group}
corresponding to ${\cal E}$. Whereas $g\mapsto U_g$ is an irreducible
projective representation of $G$, this yields an irreducible {\em
ordinary} representation of $\hat{G}$. It is a well-known fact that
$\hat{G}$ is a central extension of $G$ (cf. \cite{Isaacs:76}):
denoting the center of $\hat{G}$ by $\zeta({\hat{G}})$ this means that
$\hat{G}/\zeta(\hat{G}) \cong G$.

Given a nice error basis $\{U_g \,|\, g\in G\}$, then the abstract
error group is isomorphic to the group generated by the matrices
$U_g$. The assumption that the factor system $\alpha$ is of finite
order ensures that the abstract error group is finite.

\begin{example}[Heisenberg group]\rm 
The discrete Fourier transform of length $d \in \N$ is the unitary
transformation defined by $\DFT_d := \frac{1}{\sqrt{d}} (\omega^{k
\cdot l})_{k, l = 0, \ldots, d-1}$, where $\omega$ denotes the
primitive $d$-th root of unity $e^{2 \pi i/d}$. Define ${\cal E}_d :=
\{ S^i T^j : i = 0, \ldots, d-1, j = 0, \ldots, d-1 \}$, where
\[
S := 
\left(\begin{array}{cllc}
0 & 1 \\
  & \ddots&\ddots \\
&& \ddots & 1\\
1 &&& 0 
\end{array}\right),\quad
T := 
\DFT_d^{-1} \cdot S \cdot \DFT_d =
\left(\begin{array}{cllc}
1 &  \\
  & \omega & \\
&& \ddots & \\
 &&& \omega^{d-1}
\end{array}\right).
\]
Here again $\omega$ is the primitive $d$-th root of unity $e^{2\pi i
/d}$. Then ${\cal E}_d$ is a nice error basis on $\C^d$ showing the
existence of nice error bases for any dimension $d\in \N$ of the
underlying system. The index group in this case is the abelian group
$G = Z_d \times Z_d$ whereas the corresponding abstract error group is
a nonabelian group isomorphic to a semi-direct product $\hat{G} \cong
(Z_d \times Z_d) \rtimes Z_d$ (the so-called Heisenberg group). The
projective representation of $G$ leading to the error basis ${\cal
E}_d$ is defined by mapping the generators of $G$ as follows: $(1,0)
\mapsto S$ and $(0,1) \mapsto T$. The identity $ST = \omega TS$ is
readily verified which shows that the commutator subgroup of $\hat{G}$
is contained in the center $\zeta(\hat{G})$. This also shows that the
factor system $\alpha$ corresponding to the projective representation
of $G$ defined ${\cal E}_d$ is given by
\[ 
\alpha( (i, j), (k, l) ) = \omega^{-jk},
\]
for all $(i,j), (k,l) \in G$. 
\end{example}

We give a brief account of some general properties of nice error bases
(see also \cite{KR:2000a}). A complete classification of abstract
error groups on $\C^d$ for $1 \leq d \leq 11$ was given in
\cite{KR:2000a}. Index groups of abstract error groups are in general
not abelian: in \cite{KR:2000a} a family of groups having nonabelian
index groups was constructed. It is known that all abstract error
groups are solvable. Moreover, it is known that all solvable groups
can occur as {\em subgroups} of index groups of nice error bases. On
the other hand, it is known that not all solvable groups can occur as
index groups.

\subsection{Averaging and Annihilation}
Using the concept of abstract error groups we describe the idea of
switching off an interaction by averaging over a group. Whereas usual
techniques are based on ordinary irreducible representations
\cite{Zanardi00}, the following lemma shows that averaging over a
projective irreducible representation also projects onto the set of
scalar matrices.

\begin{Lemma}\label{averageProjective}
Let $M\in\mathbb{C}^{d\times d}$, $G$ be a finite group, and $R : g
\mapsto U_g \in {\cal U}(d)$ an irreducible projective representation
of $G$. Then the following equation holds:
\[
\frac{1}{|G|}\sum_{g\in G} U_g^\dagger M U_g = \frac{\mathrm{tr}(M)}{d}\, \onemat
\]
\end{Lemma}
{\bf Proof:} We have seen that each projective representation of an
index group $G$ with associated factor system $\alpha$ gives rise to
an ordinary representation of the corresponding abstract error group
$\hat{G}$ and that $\hat{G}$ is a central extension of $G$. It follows
that $\{U_g : g \in G\}$ is a set of coset representatives for
$\zeta(\hat{G})$ in $\hat{G}$, i.\,e.,
\[
\hat{G} = \bigcup_{g \in G} \zeta(\hat{G}) U_g.
\]
Each element $\sigma \in \hat{G}$ has a unique factorization of the
form $\sigma = z g$ where $z \in \zeta(\hat{G})$ and $g \in G$. From
Schur's Lemma (cf. \cite[Section 2.2]{Serre:77}) follows that for
$M\in\mathbb{C}^{d\times d}$, $\hat{G}$ a finite group, and $R :
\sigma \mapsto U_\sigma \in {\cal U}(d)$ an irreducible (ordinary)
representation of $\hat{G}$ the following identity holds:
\begin{equation}\label{schur}
\frac{1}{|\hat{G}|}\sum_{\sigma\in \hat{G}} U_\sigma^\dagger M U_\sigma =
\frac{\mathrm{tr}(M)}{d}\, \onemat.
\end{equation}
Using this we obtain
\begin{eqnarray*}
\frac{1}{|G|}\sum_{g \in G} U_g^\dagger M U_g & = &
\frac{1}{|G|}\frac{1}{|\zeta(\hat{G})|} 
\sum_{g \in G} \sum_{z\in\zeta({\hat{G}})} U_g^\dagger U_z^\dagger M U_z U_g \\
& = & \frac{1}{|\hat{G}|}\sum_{\sigma\in\hat{G}} U_\sigma^\dagger M U_\sigma \\
& = & \frac{{\rm tr}(M)}{d}\, \onemat,
\end{eqnarray*}
where the last line is due to Schur's Lemma (\ref{schur}) for ordinary
representations. \qed

\subsection{Decoupling}
We consider a bipartite quantum system (e.\,g.\ a system coupled to a
bath) living on the joint Hilbert space ${\mathcal H}_S\otimes
{\mathcal H}_B$. Here ${\mathcal H}_S$ and ${\mathcal H}_B$ denote the
Hilbert spaces of $S$ and $B$ respectively. Let $su(\mathcal{H})$
denote the Lie algebra of traceless self-adjoint matrices acting on
the Hilbert space $\mathcal{H}$. The Hamiltonian can be written as
\begin{equation}
H=H_S\otimes \onemat_B+\onemat_S\otimes H_B+H_{SB}\,,
\end{equation}
where $H_S\in su({\mathcal H}_S)$ is the free system Hamiltonian,
$H_B\in su({\mathcal H}_B)$ is the free bath Hamiltonian, and $H_{SB}$
describes the coupling between the system and the bath, i.\,e.\
$H_{SB}=\sum_j A_j\otimes B_j$ with $A_j\in su({\mathcal H}_S)$ and
$B_j\in su({\mathcal H}_B)$.  In order to protect the evolution of $S$
against the effect of the interaction $H_{SB}$ we seek a cyclic pulse
sequence as a suitable decoupling interaction.

By applying the pulse sequence of an annihilator we get
\begin{equation}
\tilde{H}= \frac{1}{|G|}\sum_{g\in G} 
(U_g^\dagger\otimes \onemat) H (U_g\otimes \onemat) =
\onemat_S\otimes H_B\,.
\end{equation}
This shows that decoupling can be achieved using an annihilator
procedure on only one of the subsystems
(cf. \cite{violaDecoupling,Zanardi00}).

\subsection{Inversion of Hamiltonians}\label{inversion}
We consider the problem to invert an arbitrary, possibly unknown
Hamiltonian, i.\,e.\ to simulate $-H$ given the Hamiltonian $H$.

We can use the following trick \cite{JBdistinguish}: by averaging over
all elements of $G$ but the identity we can invert the Hamiltonian
\begin{equation}
\sum_{g\in G\setminus\{1\}} U_g^\dagger H U_g = -H\,
\end{equation}
because of Lemma~\ref{averageProjective} (note that $H$ is
traceless). The resulting time overhead is $|G|-1=d^2-1$ and the
complexity is $d^2-1$. This can be seen as a generalization of the
refocussing technique used in NMR.
In general, the inverted time evolution will be  slower
than the original one:

\begin{Lemma}[Lower bound on inverting]
Let $r$ be the greatest eigenvalue and let $q$ be the smallest
eigenvalue of $H$. Then $\tau\ge\frac{r}{-q}$ is a lower bound on the
overhead for simulating $-H$ by $H$.
\end{Lemma}
{\bf Proof:}
Write $-H$ as a positive linear combination of conjugates of $H$ as 
in Definition~\ref{simulating}.
Let $\lambda_{\min}(A)$ be the smallest eigenvalue of an operator 
$A$. Then we have 
\[
-r=\lambda_{\min} (-H)=\lambda_{\min} \Big(\sum_i \tau_i U^\dagger_i H
U_i \Big) 
\geq \tau \lambda_{\min} (H)= \tau q.
\]
The inequality is due to
$\lambda_{\min}(A+B)\ge\lambda_{\min}(A)+\lambda_{\min}(B)$ (see
\cite{bha}, Theorem~III.2) for the sum of two Hermitian matrices $A$
and $B$.  Since $q$ is negative it follows that $-r/q \leq \tau$.  \qed

For the Hamiltonian $H=\mathrm{diag}(d-1,-1,\ldots,-1)$ the overhead
is at least $d-1$. Therefore a lower bound on time overhead for
inverting an unknown Hamiltonian is $d-1$.

%
%

\section{Universal transformation of Hamiltonians}\label{transformer}

In Section~\ref{annihilator} we have given a necessary and sufficient
condition on the minimal set of available control operations in order
to enable inversion and cancelling of Hamiltonians. If we want to
simulate an arbitrary Hamiltonian by any other this
condition is not sufficient. This can be seen by the following
example. Assume that the only control operations on $\C^2$ are given
by the Pauli-matrices (in their role as unitary operators). If the
Hamiltonian $H:=\sigma_z$ is given, conjugation of $H$ by a
Pauli-matrix can only lead to either $H$ or $-H$. All the Hamiltonians
which can be obtained as average Hamiltonians are scalar multiples of
$H$. Hence one cannot simulate e.\,g.\ $\sigma_x$. The following concept
will be useful in order to find groups which enable universal
simulation.

\begin{Definition}[Transformer]\label{transformerdef}
A subgroup $\mathcal{T}$ of $SU(d)$ is called a universal transformer of 
Hamiltonians iff every $\R$-linear map $L$ on $su(d)$  (i.\,e.,
the set of self-adjoint traceless operators) can be written as
\[
L(A)=\sum_j p_j U_j^\dagger A U_j
\]
with positive real numbers $p_j$ and $U_j \in \mathcal{T}$.
\end{Definition}

The physical of this is that a transformer allows to simulate the
Hamiltonian $L(H)$ if the {\it unknown} Hamiltonian $H$ is present.
In \cite{JBdistinguish} it has been shown that $SU(d)$ is a
transformer for every dimension $d$.

\begin{Observation}
In particular, a transformer is able to simulate an arbitrary
Hamiltonian $\tilde{H} \in su(d)$ by an arbitrary Hamiltonian $H\in
su(d)$.
\end{Observation}

Remarkably, the condition for a finite group to be a transformer can
be characterized in terms of irreducibility of certain
representations. In contrast to the condition for an annihilator, it
refers to the adjoint action on the set of operators instead of the
underlying Hilbert space.

\begin{Definition}[Adjoint action]
Let $G$ be a finite group and $\varphi$ a unitary
representation of degree $d$, i.\,e., $\varphi$ operates on $V =
\C^d$. We define a linear representation $\varphi_{{\rm ad}}$ on $V
\otimes V$ by $\varphi_{{\rm ad}}(g) := \overline{\varphi(g)} \otimes
\varphi(g)$ for all $g \in G$, where $\overline{U}$ denotes complex
conjugation of a matrix $U$. We call $\varphi_{{\rm ad}}$ the {\em
adjoint action} of $\varphi$. Note that this action can be identified
with the action of $G$ on matrices via conjugation $g \mapsto (M
\mapsto \varphi(g)^\dagger M \varphi(g))$.
\end{Definition}

In the following we make use of the fact that the algebra generated by
the images of an irreducible $m$-dimensi\-onal representation
$\vartheta$ of a finite group $G$ over the complex numbers is equal to
the full matrix algebra $\C^{m\times m}$. We cite the corresponding
theorem from \cite[Theorem 9.2]{Isaacs:76}. Recall that a
representation $\vartheta$ defined over a field $F$ is called {\em
absolutely irreducible} if it remains irreducible when considered over
an extension field $E/F$.

\begin{Theorem}\label{matrices}
Let $\vartheta$ be an absolutely irreducible representation of a finite
group $G$ which has degree $m$ and is defined over the field $F$. Then
\[
\Big\{ \sum_{g \in G} \alpha_g \vartheta(g) : \alpha_g \in
F \Big\} = F^{m \times m}.
\]
In particular for any $m$-dimensional irreducible representation over
the field $\C$ of complex numbers the vector space generated by the
images equals $\C^{m \times m}$.
\end{Theorem}

We now have the necessary prerequisites to characterize finite transformers. 

\begin{Theorem}[Characterization of finite transformers]\label{characterization}\ \\ A finite group
$\mathcal{T}\leq SU(d)$ is a transformer if and only if the adjoint
representation $\vartheta$  given by
\[
\vartheta(U) :=(A \mapsto U^\dagger A U)
\]
with $U\in\mathcal{T}$ acts irreducibly on $sl(d)=su(d) + i\,\, su(d)$,
i.\,e., the space of traceless operators.
\end{Theorem}
{\bf Proof:} $(\Leftarrow )$ Let $L$ be a given $\R$-linear map on
$su(d)$ and assume that the adjoint action of $\mathcal{T}$ is
irreducible on $sl(d)$ and denote this representation by
$\vartheta$. From Theorem \ref{matrices} follows that the complex
linear span of the images of $\vartheta$ is the full matrix algebra
acting on $sl(d)$.  Hence, the mapping $L$ can be written as a complex
linear combination of the form $L : A \mapsto \sum_i q_i U_i^\dagger A
U_i$. We now show that the coefficients can be chosen to be real:
since $U_i^\dagger A U_i$ is self-adjoint for all $i$ we have
$L(A)=L(A)^\dagger=\sum_i \overline{q_i} U_i^\dagger A U_i$. Therefore
we can write $L$ in the form $L : A \mapsto \sum_i p_i
U_i^\dagger A U_i$ with coefficients $p_i=\frac{1}{2}(q_i +
\overline{q}_i)\in\mathbb{R}$.  Using the inversion scheme of
Section~\ref{inversion} we can chose the coefficients $p_i$ to be
positive real numbers.

$(\Rightarrow )$ Assume that every $\R$-linear map on $su(d)$ can be
implemented in the sense of Definition~\ref{transformerdef} using
$\mathcal{T}$. Let $\cM$ be the complex linear span of the maps
$\vartheta (U)$ with $U \in \cT$.  The idea is to show that any $F\in
sl(d)$, $F\not=0$ can be mapped to any other $\tilde{F}\in sl(d)$ by a
map $T\in\cM$. This in turn shows that the adjoint action is
irreducible since there cannot be a nontrivial invariant subspace. To
construct $T$ proceed as follows. Let $F=H_1 + i H_2$ with $H_1, H_2
\in su(d)$.  Assume w.l.o.g. that $H_1\neq 0$, otherwise multiply $F$
by $-i$.  The set $\cM$ contains maps $L_1$ and $L_2$ with
$L_1(H_1)=\tilde{H}_1$, $L_1(H_2)=\lambda \tilde{H}_1$ and
$L_2(H_1)=\tilde{H}_2$, $L_2(H_2)=\mu \tilde{H}_2$ with $\mu,\lambda
\in \R$. Then $T:=L_1/(1+\lambda) + i\,L_2/(1+\mu)$ is the desired
map.  \qed

\subsection{Finite transformers}
We derive a necessary and sufficient condition for a finite group to
be a transformer group in the sense of Definition
\ref{transformerdef}. Theorem~\ref{characterization} shows that the
problem to construct a finite transformer group is to find for given
dimension $d > 1$ a finite group $G$ and an irreducible (unitary)
representation $\varphi$ of $G$ such that the adjoint action becomes
irreducible if we split off the trivial representation $\onemat$ of
$G$. The trivial representation is always contained in $\varphi_{{\rm
ad}}$ since the one-dimensional space corresponding to the linear span
of the identity matrix remains invariant, i.\,e., $\varphi_{{\rm ad}}
= \onemat \oplus \pi$ for some representation $\pi$ of $G$. Abusing
the notation we will write $\varphi_{{\rm ad}} - \onemat$ to denote
the summand $\pi$ in this decomposition.

Once we have found a suitable pair $(G, \varphi)$ with ${\rm
deg}(\varphi) = d$ this yields a transformer group as in Definition
\ref{transformerdef}. For basic results concerning representation
theory of finite groups we refer the reader to \cite{Isaacs:76}.

\begin{example}\label{twodimtransformer}
We examine the case of a two-dimensional system, i.\,e.,
$d=2$. Starting from the Pauli matrices 
\[
\sigma_x := \left(\begin{array}{cc} 0 & 1 \\ 1 & 0 \end{array} \right),\;
\sigma_y := \left(\begin{array}{rr} 0 & -i \\ i & 0 \end{array} \right),\;
\sigma_z := \left(\begin{array}{rr} 1 & 0 \\ 0 & -1 \end{array} \right).
\]
we first note that the group $\langle i\cdot \sigma_x, 
i \cdot \sigma_y, i \cdot \sigma_z \rangle$ is isomorphic to the
quaternion group $Q_8$ of order $8$. This group has an (outer)
automorphism of order $3$ which permutes the Pauli matrices cyclically.
This automorphism is given by the matrix 
\[
R :=
\frac{i-1}{2}\left(\begin{array}{rr} i & i \\ -1 & 1
\end{array} \right).
\]
Setting $s_k := i \sigma_k$ for $k \in \{x, y, z\}$ the automorphism
is given by $R^{-1}s_x R=s_y$, $R^{-1}s_y R=s_z$, and $R^{-1}s_z
R=s_x$.  The group generated by the $s_k$ and $R$ is isomorphic to
$\SL(2,\F_3)$, i.\,e., the group of $2 \times 2$ matrices over the
finite field $\F_3$ which have determinant $1$. Let $\varphi$ be the
(natural) representation of the matrix group given by $\langle s_x,
s_y, s_z, R \rangle$. Then the $24$ matrices in the image of $\varphi$
form a faithful irreducible representation of $\SL(2,\F_3)$. Choosing
the basis $\{ s_x, s_y, s_z \}$ of $sl(2)$ we see that the images of
$\varphi_{{\rm ad}} - \onemat$ are given explicitly by $s_x \mapsto
{\rm diag}(1,-1-1)$, $s_y \mapsto {\rm diag}(-1,1,-1)$, $s_z \mapsto
{\rm diag}(-1,-1,1)$, and $R$ maps to the permutation matrix
corresponding to the $3$-cycle $(1,2,3)$. It is readily verified that
this is an irreducible representation.
\end{example}

Let $G$ be a finite group having an irreducible representation
$\varphi$ such that the images of $\varphi$ are a transformer in the
sense of Definition \ref{transformerdef}. Then necessarily $\varphi$
must be nonmonomial\footnote{A representation is called {\em monomial}
if all representing matrices have the property to contain precisely
one non-vanishing entry in each row and each column.} for otherwise
the set of diagonal matrices would be an invariant subspace under the
action of $\varphi_{{\rm ad}}-\onemat$. Note that in fact the group
$\SL(2,\F_3)$ is the smallest group which is not an ${\cal M}$-group,
i.\,e., $\SL(2,\F_3)$ has representations which are not equivalent to
monomial ones. Therefore a necessary condition for $\varphi$ to be a
transformer has been found.

There is a necessary and sufficient characterization of transformer
groups which can be verified from the character table alone. Recall
that the character $\chi$ of a representation $\varphi$ is defined by
$\chi(g) := {\rm tr}(\varphi(g))$ and that a character is called
irreducible iff the corresponding representation is irreducible.

\begin{Theorem}
Let $G$ be a finite group and $\chi$ be an irreducible character of
$G$ with corresponding representation $\varphi$. Then $\chi$
corresponds to a universal transformer if and only if the following
identity holds:
\[ \sum_{g \in G} |\chi(g)|^4 = 2 |G|.
\]
\end{Theorem}
{\bf Proof:} The representation $\varphi_{{\rm ad}}-\onemat$ has
character values $|\chi(g)|^2-1$ for all $g \in G$ since ${\rm
tr}(\varphi_{{\rm ad}}(g)) = \overline{\chi(g)}\chi(g)$. Recall that
the vector space of class functions on $G$ has a normalized scalar
product given by
\[ 
\langle \chi_1 | \chi_2 \rangle = 
\frac{1}{|G|} \sum_{g \in G} \chi_1(g) \chi_2(g^{-1})
\]
for characters $\chi_1$, $\chi_2$ of $G$. A character $\chi$ is
irreducible iff $\langle \chi | \chi \rangle = 1$. Computing the
latter scalar product of the character corresponding to $\varphi_{{\rm
ad}}-\onemat$ we obtain
\begin{eqnarray*}
\frac{1}{|G|} \sum_{g \in G} (|\chi(g)|^2-1) \cdot (\overline{|\chi(g)|^2-1})
& = & \frac{1}{|G|} \sum_{g \in G} |\chi(g)|^4 - \frac{2}{|G|}
\sum_{g \in G} |\chi(g)|^2 + 1 \\
& = & \frac{1}{|G|} \sum_{g \in G} |\chi(g)|^4 - 1
\end{eqnarray*}
On the other hand this scalar product is equal to $1$ due to the
irreducibility of $\varphi_{{\rm ad}}-\onemat$. Rearranging terms and
clearing denominators yields the claimed statement.  \qed

In the following we present a transformer for a three dimensional
system. The minimal group having a representation $\varphi$ for which
$\varphi_{{\rm ad}}-\onemat$ is irreducible is the linear group
$\GL(3,\F_2)$ of invertible $3 \times 3$ matrices over the field
$\F_2$. This is a simple group of order $168$. As generators of this
group we choose the matrices
\[
x := 
\left(\begin{array}{ccc}
1 & 1 & 0 \\
0 & 1 & 0 \\
0 & 0 & 1
\end{array}\right), 
\;
y := 
\left(\begin{array}{ccc}
1 & 1 & 1 \\
1 & 1 & 0 \\
1 & 0 & 0
\end{array}\right),
\]
where $x$ is an element of order $2$ and the order of $y$ is $7$. The
group $\GL(3,\F_2)$ has a three-dimensional irreducible representation
$\varphi$ over the complex numbers which on the generators $x$ and $y$
is given by the following assignments:
\[
\varphi(x) := 
\frac{2}{\sqrt{7}} 
\left(\begin{array}{rrr}
\cos(\frac{5\pi}{14}) & - \zeta_7^3 \cos(\frac{\pi}{14}) & -\zeta_7^2 
\cos(\frac{3 \pi}{14}) \\[1ex]
-\zeta_7^4 \cos(\frac{\pi}{14}) & -\cos(\frac{3\pi}{14}) & \zeta_7^6
\cos(\frac{5 \pi}{14}) \\[1ex]
-\zeta_7^5\cos(\frac{3\pi}{14}) & \zeta_7 \cos(\frac{5\pi}{14}) & 
-\cos(\frac{\pi}{14}) 
\end{array}\right), 
\;
\varphi(y) := 
\left(\begin{array}{rrr}
\zeta_7 & \cdot & \cdot \\
\cdot & \zeta_7^2 & \cdot \\
\cdot & \cdot & \zeta_7^4
\end{array}\right).
\]
Here $\cdot$ is an abbreviation for $0$ and $\zeta_7$ denotes the
primitive $7$-th root of unity $e^{2\pi i/7}$. The character of the
representation $\varphi$ takes the values 
\[
 3,\; -1,\; 1,\; 0,\; \zeta_7+\zeta_7^2+\zeta_7^4,\;  \zeta_7^3+\zeta_7^5+\zeta_7^6 
\]
on the conjugacy classes of $\GL(3,\F_2)$. Consulting the character
table of $\GL(3,\F_2)$ in the Atlas \cite[p.~3]{Atlas:85} we find that
$\varphi$ is irreducible. The representation $\varphi_{{\rm
ad}}-\onemat$ has character values
\[
 8,\; 0,\; 0,\; -1,\; 1,\; 1
\]
from which follows that it is also irreducible, again by checking the
character table of $\GL(3,\F_2)$. Overall we obtain that the
representation $\varphi$ of $\GL(3,\F_2)$ yields a transformer of size
$168$. Using the Neub{\"u}ser catalogue used in MAGMA and GAP,
cf.~\cite{magma,gap} we performed an exhaustive search over all groups
of smaller sizes which has shown that this indeed is the minimal
possible group size.

In Tabular \ref{transformerTab} we summarize the results of this
search. Groups of sizes up to $255$ have been considered. The number
in the Neub{\"u}ser catalogue is given such that for instance the first
row of this table corresponds to the group (in GAP syntax) {\tt
SmallGroup(24,3)} which has been studied in Example
\ref{twodimtransformer}. Note that we only give transformer groups
which act faithfully.

\begin{table}
\begin{center}
\begin{tabular}{|c|c|c|}
\hline
Group size & Numbers in library & Dimension \\
\hline\hline
24 & 3 & 2 \\ 
48 & 28, 29, 33 & 2 \\ 
72 & 3, 25 & 2 \\ 
96 & 67, 74, 192 & 2 \\ 
120 & 5 & 2 \\ 
144 & 36, 121, 122, 157 & 2 \\ 
168 & 22 & 2 \\ 
168 & 42 & 3 \\ 
192 & 187, 204, 963 & 2 \\ 
216 & 3, 38 & 2 \\
216 & 88 & 3\\
240 & 93, 102, 103, 154 & 2 \\
\hline
\end{tabular}
\caption{\label{transformerTab} Transformer groups of small sizes}
\end{center}
\end{table}

\subsection{Lower bound on the overhead}
In the following we derive a lower bound on the time overhead for
simulating a Hamiltonian using an arbitrary transformer. We need
some results on majorization and doubly stochastic matrices
(cf. \cite{vidalmaj} for a summary).  Let $x=(x_1,\ldots,x_d)$ and
$y=(y_1,\ldots,y_d)$ be two $d$-dimensional real vectors. We introduce
the notation $\downarrow$ to denote the components of a vector
rearranged into non-increasing order, so
$x^\downarrow=(x_1^\downarrow,\ldots,x_d^\downarrow)$, where
$(x_1^\downarrow\ge x_2^\downarrow\ge\ldots\ge x_d^\downarrow)$. We
say that $x$ is majorized by $y$ and write $x\prec y$, if
$$ \sum_{j=1}^k x_j^\downarrow\le \sum_{j=1}^k y_j^\downarrow\,, $$
for $k=1,\ldots,d-1$, and with equality when $k=d$ \cite{bha}.

Let $\mathrm{Spec}(X)$ denote the spectrum of the hermitian matrix
$X$, i.\,e.\ the vector of eigenvalues.
Ky Fan's maximum principle
gives rise to a useful constraint on the eigenvalues of a sum of two
Hermitian matrices $C:=A+B$, that
\begin{equation}\label{majorsum}
 \mathrm{Spec}(A+B)\prec \mathrm{Spec} (A)+ \mathrm{Spec}(B)\,.
\end{equation}
This permits us to derive a lower bound on the simulation overhead.
\begin{Lemma}[Lower bound]
A lower bound on the overhead of simulating $\tilde{H}$ by $H$ is
given by the minimal $\tau$ such that
\begin{equation}
\mathrm{Spec}(\tilde{H})\prec\tau\mathrm{Spec}(H)\,.
\end{equation}
\end{Lemma}
{\bf Proof:} This follows from Definition~\ref{simulating} and
inequality (\ref{majorsum}). \qed

We now consider the question when this lower bound can be attained.
Let $H=\sum_{i=1}^d \lambda_i |i\rangle\langle i|$ and
$\tilde{H}=\sum_{i=1}^d\mu_i |i\rangle\langle i|$ where $|i\rangle$ is
a basis of eigenvectors. Let $\tau$ be minimal such that
$\vec{\mu}\prec\tau\vec{\lambda}$.  We set $d':=d-1$. We have
$(\mu_1,\ldots,\mu_{d'})\prec (\tau\lambda_1,\ldots,\tau\lambda_{d'})$
since $\sum_i \mu_i=\sum_i \lambda_i = 0$ (the Hamiltonians are
traceless). This is equivalent to the existence of a doubly
stochastic matrix $D$ with
$D(\lambda_1,\ldots,\lambda_{d-1})^T=(\mu_1,\ldots,\mu_{d-1})^T$. By
Birkhoff's theorem we can decompose a doubly stochastic matrix as a
convex sum of permutations, i.\,e.\
\begin{equation} 
D=\sum_{\sigma\in\Sigma} p_{\sigma} U_\sigma
\end{equation} 
where $U_\sigma$ is the permutation matrix associated to $\sigma$,
i.\,e.\ maps the basis vector $|i\rangle$ to $|\sigma(i)\rangle$, and
and $\Sigma$ is a subset of the symmetric group $S_{d'}$. The
$d'\times d'$ doubly stochastic matrices form a
$({d'}^2-2d'+1)$-dimensional convex set. The extreme points are the
permutation matrices. Carath\'{e}odory's theorem guarantees that a point
in a $m$-dimensional compact convex set may be expressed as a convex
combination of at most $m+1$ extremal points of that set. Therefore
every doubly stochastic matrix can be written as a convex combination
of at most ${d'}^2-2d'+2=(d-2)^2+1$ permutations.

We view the matrices $U_\sigma$ as $d\times d$ matrices that fix
the basis vector $|d\rangle$. We have 
\begin{equation}
\tilde{H}=\tau\sum_{\sigma\in\Sigma} p_{\sigma} U^{\dagger}_{\sigma} H
U_{\sigma} 
\end{equation} 
The lower bound can be attained in particular if the transformer contains
the matrices permuting the eigenvectors of $H$ and and the matrix realizing
the base change between the eigenvector basis of $H$ and $\tilde{H}$.
 
Let $H$ be the system Hamiltonian. Let $\sigma$ be the cyclic shift,
i.\,e.\ $\sigma(i)=i+1\mod d$ and $U_\sigma$ the corresponding
matrix. Than we can switch off the Hamiltonian 
\begin{equation}
\sum_{j=0}^{d-1} \frac{1}{d} U_{\sigma^j}^\dagger H U_{\sigma^j} = 0
\end{equation} 
with complexity at most $d$ provided that we can perform the shift
with the transformer. In this case we can also invert it with
complexity and overhead at most $d-1$. However, when the Hamiltonian
is not known we need at least $d^2$ operations to switch it off.

\section{Simulation of bipartite Hamiltonians}\label{bipartite}
The following theorem shows that all bipartite Hamiltonians
can be simulated by any Hamiltonian (with non-trivial coupling and non-trivial
local terms) provided that the set
of available unitary transformations contains a transformer
for each of the subsystems. Let $B=\{\sigma_\alpha\mid
\alpha=1,\ldots,d^2-1\}$ be a basis of $su(d)$.

\begin{Theorem}
Let an arbitrary interaction
\[
H:=\sum_{\alpha\beta} J_{\alpha \beta} \sigma_\alpha \otimes \sigma_\beta +
a\otimes\onemat + \onemat\otimes b
\]
be given with $a,b\in su(d)$ with $a,b \neq 0$. Let $\cT_1$ and
$\cT_2$ be transformers of the left and the right subsystem,
respectively. Assume that it is possible to implement all unitary
transformations of the form $U\otimes V$ with $U\in \cT_1$ and $V\in
\cT_2$. Then $H$ can be used for simulating any arbitrary $\tilde{H}$
with
$
\tilde{H}:=\sum_{\alpha\beta}
\tilde{J}_{\alpha \beta}\sigma_\alpha\otimes\sigma_\beta +
\tilde{a}\otimes\onemat + \onemat\otimes\tilde{b}
$
i.\,e., there are positive numbers $\tau_j$, $U_j\in\tau_1$ and
$V_j\in\tau_2$ such that
\begin{equation}\label{Tilde}
\tilde{H}=\sum_j \tau_j (U_j^\dagger \otimes V_j^\dagger) H (U_j
\otimes V_j)\,.
\end{equation}
\end{Theorem}
{\bf Proof:} We first consider the case that the local terms are all zero.
Write $H$ in the form 
\[
H= \sum_j A_j \otimes B_j \, ,
\]
where $A_j$ and $B_j$ are elements of $su(d)$ and all $B_j$ are
linearly independent and all $A_j$ are nonzero. Then $H$ can be
transformed into any interaction of the form $C\otimes D$ with
arbitrary $C,D \in su(d)$. This can be done by choosing $\R$-linear
maps $L_1$ and $L_2$ on $su(d)$ with $L_1(A_1)=C$, $L_2(B_1)=D$, and
$L_2(B_j)=0$ for $j\neq 1$.  Since $\cT_1$ and $\cT_2$ are universal
transformers one can find positive numbers $d_j$ and $f_j$ and unitary
transformations $U_j \in \cT_1$ and $V_j \in \cT_2$ such that
\[
L_1(\cdot) = \sum_j d_j U_j^\dagger\cdot U_j \,\,\hbox{ and }\,\, 
L_2(\cdot) = \sum_j f_j V_j^\dagger\cdot V_j\,.
\]
Hence we obtain
\[
\sum_{ij} d_if_j (U_i^\dagger\otimes V_j^\dagger) H (U_i\otimes V_j) = 
C\otimes D \,.
\]
This proves that we can simulate each tensor product operator
$C\otimes D$.  By setting $C:=\tilde{J}_{\alpha\beta} \sigma_\alpha$
and $D:=\sigma_\beta$ we can simulate each term in eq.~(\ref{Tilde}).
Hence it is possible to simulate $\tilde{H}$. Let $H$ contain local
terms. Starting from $H$ we can simulate the Hamiltonian
\[
H'=\sum_{\alpha\beta}\tilde{J}_{\alpha\beta}\sigma_\alpha\otimes\sigma_\beta
+a'\otimes\onemat + \onemat\otimes b'\,,
\]
that coincides with the desired Hamiltonian $\tilde{H}$ except for the
local terms. Starting from $H$ we can simulate the Hamiltonian
$(\tilde{a}-a')\otimes\onemat$ by applying an annihilator on the right
and suitable transformations on the left.  Finally, we use a similar
scheme for $\onemat\otimes (\tilde{b}-b')$.  
\qed

\section{Conclusions}
We have shown that there are finite groups of unitary control
operations which allow to simulate an arbitrary Hamiltonian
$\tilde{H}$ by another arbitrary $H$, i.\,e., a system with
Hamiltonian $H$ can be driven to evolve as if its Hamiltonian were
$\tilde{H}$. This can be accomplished using fast sequences of control
operations interspersing the natural time evolution. We even found
finite groups which allow to solve the following more general control
problem: for every linear trace preserving map $L$ on the set of
self-adjoint operators the system can be made to evolve according to
the Hamiltonian $L(H)$ although its true {\it unknown} Hamiltonian is
$H$. We have called such groups {\it transformer groups} and showed
that a finite group $G$ has this property if and only if its adjoint
action on the set of traceless operators is irreducible.  We have
characterized finite groups $G$ with this property using characters of
representations of $G$ on the Hilbert space. This criterion allows to
perform an exhaustive search over groups of small order (up to $255$).
We found transformer groups for two and three dimensional quantum
systems. It remains an open problem to construct finite transformer
groups for all dimensions. In bipartite systems, every non-trivial
interaction can simulate any other provided that transformer groups on
each subsystem can be implemented.

\section*{Acknowledgments}
We would like to thank Micha{\l} Horodecki for helpful comments. This
work has been supported by the European Community through grant
IST-1999-10596 (Q-ACTA) and the DFG project {\em Komplexit{\"a}t und
Energie}.

%
%

\bibliographystyle{plain} 
\bibliography{transformer}

\end{document}